\documentclass[slac_one]{revtex4}
\usepackage{graphicx}
\usepackage{fancyhdr}
\pdfoutput=1

\begin{document}

\title{Recent Progress in Formal Theory} 

%

\author{Joseph Polchinski}
\affiliation{KITP, UCSB, Santa Barbara, CA 93106-4030, USA}

\begin{abstract}
This is a summary talk covering recent progress in perturbative methods for gauge theories and gravity, applications of AdS/CFT duality, vacuum energy, and string theory models of particle physics and cosmology.
\end{abstract}

\maketitle

\thispagestyle{fancy}


\section{I DIDN'T CHOOSE THE TITLE} 

The organizers chose it.  I'm not sure what `formal theory' is supposed to cover, but they made the useful suggestion that my job is to summarize the parallel talks.  There were 21 talks,  covering a good range of interesting topics.  With just a little squeezing the talks fall into four broad categories:
perturbative methods in gauge theory and gravity, applications of AdS/CFT duality, vacuum energy, and string models of particle physics and cosmology.  For reasons of space, I will cite the parallel talks rather than the original papers; please see those talks for detailed references.

\section{PERTURBATIVE METHODS}

Feynman graphs are central to quantum field theory, but it is now a familiar idea, going back more than twenty years to Parke and Taylor, that even in perturbation theory there can be better ways to organize things.  In particular there are major simplifications of the amplitudes that are not evident in the graphical expansion.  For example, the $n$-gluon amplitude with maximally helicity violating (MHV) polarization would be given by a sum of thousands of terms from Feynman graphs, yet the final answer collapses into a single term.  And progress continues, as reviewed by Britto~\cite{Britto}.  For the next-to-MHV amplitude, the original spinor helicity methods give an expression that fits on a single slide.  With twistor methods, this is reduced to a couple of lines, and with the newest residue methods one obtains an expression nearly as compact as for the MHV amplitudes.

These techniques have wide applications, to Standard Model processes, to model systems like ${\cal N}=4$ supersymmetry, and to (supergravity).  The Standard Model applications were discussed in Zanderighi's plenary talk~\cite{Zanderighi}, so I will focus on the more formal directions.\footnote{I should mention also the parallel talk by Marquard~\cite{Marquard}, which discussed simplifications along a different axis: using integration by parts to reduce large numbers of loop integrals to a much smaller standard set, in an efficient way.  Applications included the low energy expansion of the photon polarization function, to four loops, and the determination of charm and bottom quark masses.}

\subsection{Gauge Theory}

A natural question is whether these ongoing simplifications are a sign of some deeper structure, like a symmetry.  In Maldacena's talk~\cite{Maldacena}, it was shown that the answer is `yes', at least in part.  To explain what he found, I need to recall the 1974 observation of 't Hooft, that to understand QCD we should think first about the limit where the number of colors $N_{\rm c}$ is taken to be large, with $\alpha_{\rm s} N_{\rm c}$ held fixed.  In this limit, only the planar Feynman graphs survive.  One still has an infinite number of graphs, each quite complicated, so this does not immediately give a solution (unlike some simpler large-$N$ systems).   Nevertheless, 't Hooft has given us a tantalizing and inspiring goal, to solve the theory in this limit.

One reason one might hope for progress is that in this limit the problem is reduced from 3+1 dimensions to 1+1 dimensions, since we can draw all of the graphs on a piece of paper, without crossings.  Theories in 1+1 dimensions are often simpler.  For example, Bethe was able to write down the exact many-body wavefunctions for certain spin systems in one spatial dimension: they reduce to the product of free wavefunctions, times a factor that depends only on the ordering of the particles along the line.  It is important that not all systems in 1+1 dimensions have a wavefunction that can be expressed in this way.  Rather, it is related to the presence of a symmetry, of a somewhat unusual type.  For most of the familiar symmetries in physics, the conserved charge is the integral of a local current.  The Bethe Ansatz charges are given as the double (or multiple) integral of a product of local currents, times a function again of the order of the currents along the line.

Such charges are evidently present in the planar limit of the ${\cal N}=4$ theory: this is known as `integrability,' and has been the subject of major research activity in the past few years.  It has been mysterious how this is related to the amplitude simplifications, but now Maldacena reports a connection: one of the observed properties of the scattering amplitudes, `dual conformal symmetry,' turns out to be a symmetry of the Bethe Ansatz type.  This may be the tip of a large and interesting iceberg.

As further progress toward realizing 't Hooft's goal for ${\cal N}=4$, Bern~\cite{Bern} discussed a conjecture for the exact MHV amplitudes: the all-loop amplitudes seem to have a simple iterative structure.  For the 2-to-2 amplitudes, this seems to correctly match on to the strong coupling calculation from the dual string theory, but for the 2-to-4 amplitudes there is a discrepancy to be understood.

\subsection{Gravity}

Newton's constant $G_N$ has units of mass-squared, like Fermi's constant $G_F$, and so the dimensionless coupling $G_N E^2$ grows with energy: the theory is nonrenormalizable.  Two natural resolutions suggest themselves.  The first is that gravity is only an effective theory,\footnote{The talk by Rahman~\cite{Rahman} studied the extension of this effective field theory point of view to theories of charged spin-two fields.}  and that new physics comes in at high energy that gives finite amplitudes.  In the case of the weak interaction, this is what happens.  The divergences of the Fermi theory were an important clue: they point almost uniquely to spontaneously broken non-Abelian gauge theory.  Similarly in string theory new physics enters at high energy.

A second possibility is that the divergences are an artifact of perturbation theory and cancel in the full theory.  The talk by Ward~\cite{Ward} presented evidence that this happens, through a resummation of perturbation theory that suppresses the graviton propagator at large momentum.  This is a technical subject that is difficult to assess, but I will raise one issue that comes to mind.  Attempts to cut off gravity covariantly often lead to difficulties because the Ward identity relates the 3-point vertex to the inverse propagator.  If the propagator is suppressed, the vertex is correspondingly enhanced so there is no net effect in loops.  This might also happen in attempts to resum consistently.

The talk by Bern~\cite{Bern} presented a third possibility, at least for the most supersymmetric gravity theory, ${\cal N}=8$.  The supersymmetry guarantees that there are no divergences at one or two loops, and it was expected by many supergravity experts that the first divergence would appear at three loops.  The calculation has now been done, and there is no three-loop divergence.  There are cancellations that may not be due to supersymmetry.  This raises the possibility that these cancellations continue to higher orders, and that the amplitudes are finite order by order.

This would be a remarkable result, whose physical interpretation is not at all clear.  The dimensionless coupling still blows up at high energy, so one is working with a perturbation theory that breaks down, and a nonperturbative completion is needed.  The cancellation doesn't fit into Wilson's way of looking at field theory: a UV fixed point would show up as a cancellation of divergences between different orders rather than their absence order-by-order.  But it is not at all clear to what order supersymmetry provides protection: Bern quotes various arguments, which predict that the first divergences might arise at three, five, six, seven, eight, or even nine loops!\footnote{The most naive supersymmetry argument gives approximately eight: each supersymmetry gives four superspace integrals in the counterterm, which are dimensionally the same as two derivatives or one power of $G_N$.  However, there is not a complete superspace formulation of the theory.}  So it is not clear how far one must calculate to be sure that the finiteness is not just a consequence of supersymmetry that will eventually fail at some order.

\section{APPLICATIONS OF GAUGE/GRAVITY DUALITY}

We discussed 't Hooft's idea that large-$N_{\rm c}$ QCD might be solvable, but he also went further and proposed that the solution might involve rewriting QCD as a string theory!  Roughly speaking, at strong coupling one might imagine that the planar Feynman graphs become dense with propagators and vertices, forming a continuous two-dimensional surface which is the world-sheet of the string.
The obvious problem is that attempts to quantize strings directly lead to gravity and extra dimensions, not to QCD.  This was resolved by Maldacena in a surprising way:\footnote{Some features of this  approach were anticipated by Polyakov.}  by putting the strings into anti-de Sitter spacetime, the boundary conditions freeze gravity, and reduce the kinematics to four dimensions.

So far, this does not apply to QCD.  Maldacena's original duality was for the same ${\cal N}=4$ theory that we discussed earlier, and indeed the string description has been useful in understanding some of the behaviors seen in perturbation theory.  The ${\cal N}=4$ theory is conformally invariant and so has no masses --- it is a theory of `unparticles,' in the current jargon.  One can then deform toward progressively less symmetric duals: the conformal invariance is useful to get oriented but is not central to the duality.  For this reason we use the more general term `gauge/gravity duality' (or `field/string duality') rather than the more specific AdS/CFT.

So far this principle has not been extended in a precise way to asymptotically free theories like QCD.  However, the ability to solve strongly coupled quantum field theories, to have a tool completely complementary to perturbation theory, is a terrific thing.  It is like the harmonic oscillator for strongly coupled gauge theories, and has had surprising even to real QCD.  This subject was beautifully reviewed in the plenary talk by Rajagopal~\cite{Rajagopal}, so I will mention just a few highlights.

One of the surprises from RHIC is that the guark-gluon plasma is strongly-coupled, behaving like a liquid rather than a gas.  We have no good tools from quantum field theory to study strongly coupled non-equilibrium physics, so the solvable ${\cal N}=4$ theory provides at least a reality check, a model system to compare with.  In fact it seems much better than that, in that its extra symmetries (supersymmetry, conformal invariance) have little consequence in the finite temperature plasma, and moreover there seem to be some universal properties of strongly coupled gauge theories that one can learn from.  Two important numbers are the free energy, which one learns is only corrected by about 25\% as compared to the free theory, and the shear viscosity to entropy ratio, where one learns that the remarkably small value of $1/4\pi$ is attained in many strongly coupled systems.  There were conjectures that this might represent a universal lower bound, but in the talk by Myers~\cite{Myers} it was shown that the model-dependent first correction can have either sign.  Thus it is a sort of ideal value, but not a bound.   Another example of the value of an exactly solvable system has been in the development of relativistic hydrodynamics.  This was reviewed in the talk by Starinets~\cite{Starinets}.  

Rajagopal gave a long list of other areas where the dual theory has given valuable insights~\cite{Rajagopal}.  I refer you to his talk for the details, and will focus on the conceptual question, ``Is RHIC really making black holes?"  The obvious answer is, ``of course not, it's just quarks and gluons,'' but I would argue that the truth is more subtle.  A duality means that we have a quantum theory that has more than one classical limit.  In the present case, one limit is gauge theory and the other is string theory.  Such a situation erases distinctions that we make with our classical experience.  The familiar example is wave/particle duality, but now we have the same thing happening between black holes and the quark-gluon plasma.

Of course, with wave/particle duality there are situations where things are clearly wave-like or clearly particle-like, but we know that we can interpolate between these.  With gauge/gravity duality, to interpolate we have to imagine varying the parameters of the theory.  In a sense we can do this, because in Randall-Sundrum models one has physics with both gauge and gravitational descriptions, and depending on the parameters the physics is closer to one or to the other.  But even in QCD, the surprise has been how successful the gravitational description is.  For the shear viscosity, and other properties, the best calculations have not had anything to do with quarks and gluons, but instead come from the propagation of gravitational fields near a black hole!

There are many condensed matter systems with `quantum critical points,' strongly coupled phases that have universal low energy behaviors but no simple description.  For example, it is believed that the high temperature superconductors exhibit this phenomenon.  There has recently been some attention to the idea that there might be model systems with gravitational duals, where one might be able to do calculations and gain insight that cannot be done in any other way.  Some condensed matter systems have effectively relativistic behavior at the critical point, but others are nonrelativistic, with space and time scaling differently.  The talk by Kachru~\cite{Kachru} described a class of dual systems with this property.

\section{VACUUM ENERGY}

It was good to see several talks about vacuum energy, because this is such an important problem.\footnote{For a more detailed discussion see Ref.~\cite{Polchinski}.}  The  vacuum is highly nontrivial --- full of Higgs fields, color fluctuations, zero point energies --- and gravity is a universal force.  So why is the vacuum energy not huge?  This question has been around since Pauli, but since the discovery of the cosmic acceleration we have two new puzzles: why is the vacuum energy not exactly zero, and why is its magnitude so similar to the matter density in the universe in the current era (cosmic coincidence).

These new questions get a lot of attention, but the original problem of 60 or 120 orders of magnitude is still there, and this is what I will focus on.  Many many solutions have been proposed over the years, but it is very difficult to get around the basic problem: the vacuum is full of stuff, and everything gravitates.  A useful litmus test for proposed solutions is to ask what cancels the electron zero point energy (ZPE).  
At the scale of the electron mass we know that quantum field theory works to high precision, and the modes from this scale and below generate a vacuum energy 30 orders of magnitude larger than that observed.  This is large enough, as Pauli noted, that the universe ``would not even reach out to the
moon''~\cite{Straumann}.

One might imagine that for some reason zero point energy does not gravitate, but any such idea will conflict with the great success of quantum field theory.  For example, in the Standard Model the top quark ZPE is an important contribution to the Higgs potential and the Higgs mass.  If this did not gravitate it would feed through to the rest of the Standard Model and spoil the universality of gravity, conflicting strongly both with theory and experiment.  As another example, the gravitation of the vacuum polarization contribution to atomic energy levels, which is a shift in the zero point energy due to the field of the atom, can be seen to high precision in the Eotvos experiment.  The talk by Milton~\cite{Milton} showed that the gravitation of Casimir energy is consistent with renormalization. 

If the vacuum energy were exactly zero, it would be natural to look for a symmetry explanation.
Supersymmetry can set the vacuum energy to zero.  Symmetry under scaling of lengths and energies  (or conformal transformations) could also have this effect.  But if these are symmetries of nature they must be broken, since the electron has nonzero mass and is not degenerate with the selectron.  Thus these ideas do not pass the litmus test, because the zero point energy depends on the electron mass and so, even if it vanishes in a symmetric state, it is induced by the breaking.

Two of the parallel talks explored symmetry solutions to the cosmological constant problem~\cite{Mannheim, Nevzorov}.  Mannheim  proposed a solution based on broken conformal symmetry, but I believe that in the end the net contribution of the electron ZPE is exactly the same is it would be without this symmetry.  Nevzorov revisits the interesting no-scale structure, where the vacuum energy in certain supergravity theories is zero even after supersymmetry breaking.  However, this is only a classical result, and it does not survive loop corrections such as the ZPE.  So these two attempts based on symmetry seem to run into the usual problems.

Besides symmetries, the other broad class of ideas is adjustment mechanisms.  In addition to the known physics there is an `adjustment sector,' whose vacuum energy is variable in some way so as to cancel that from the rest of the theory.  The adjustment sector must have a large number of stable states (vacua) with different values of the energy density.   But how can such an adjustment sector arise from fundamental physics?  In fact, string theory seems to provide a realization of this: by varying the shape of the compact dimensions, and the configurations of branes and fluxes, there is the potential to have an enormous number of long-lived vacua, with finely spaced values of the vacuum energy.

\section{STRING THEORY MODELS OF PARTICLE PHYSICS AND COSMOLOGY}

\subsection{Particle Physics}

As discussed above, we have a clue from cosmology that the fundamental theory must have many vacua, with adjustable vacuum energy.  As model builders, it would be nice if only the vacuum energy adjusted while all the other observable parameters descended from the underlying theory in a fixed way.  Unfortunately there is no particular reason that nature should be so nice to us, and this does not seem to be what happens in string theory: the other constants also vary with the geometry of the compact dimensions.

This of course makes it challenging to connect the underlying theory to observation, but the parallel session speakers were undaunted.  Even very large numbers of vacua need not be an obstacle: sometimes, as in statistical mechanics, one can make predictions precisely because the number of available states is very large.  The challenge for now is to map out this space of states, and figure out the rules by which it is populated.

Gmeiner~\cite{Gmeiner} considered a large class of vacua based on intersecting D6-branes and O6-planes.  This is a broad framework for constructing Standard Model-like vacua, and general features may emerge that depend on the brane configuration but are insensitive to other details of the compactification.  He presents statistics for various characteristics such as the number of generations.  In contrast to earlier work along the same lines, the number three does not seem to be disfavored.

Ratz~\cite{Ratz} explored heterotic orbifolds with `local grand unification,' where the grand unified (GUT) symmetry is broken locally in the compact dimensions.  In this way the successful predictions of GUTs can survive, such as the unification of couplings, while some of the problems are absent, notably the color triplet partners of the Higgs that would induce rapid proton decay.  This was a very optimistic talk: putting in only a small number of ingredients, such as a unification structure and $R$ symmetry, the authors found interesting results for the flavor structure, neutrino see-saw, supersymmetry breaking, and the $\mu$-problem.  One of the big questions for the LHC is whether we can find more evidence for GUTs.  The prediction of $\sin^2 \theta_{\rm w}$ is striking, but with just one number it may still be an accident.

Camara~\cite{Camara}, Krefl~\cite{Krefl}, and Weigand~\cite{Weigand} all discussed the technology of the calculation of string loop and instanton corrections, which may be important for supersymmetry breaking and mediation, the $\mu$ term, and neutrino and other fermion masses.  Supersymmetry breaking and its mediation is one place where observable physics may connect closely to the short distance theory, and in particular to the geometry of the compact dimensions.  Seiberg~\cite{Seiberg} discussed gauge mediation in a model-independent way.  There are several generic predictions: sfermion degeneracy and absence of flavor changing neutral currents, general mass relations ${\rm Tr} \, m_{\tilde f}^2 (B-L) =
{\rm Tr}\, m_{\tilde f}^2 (B-L) = 0$, small $A$ terms, problems with $\mu$ and $\mu B$, and the gravitino as the lightest supersymmetric partner.  Other predictions, such as gaugino and mass sfermion relations, are model dependent.  Weigand~\cite{Weigand} presented a string theory realization of gauge mediation.

\subsection{Cosmology}

Early cosmology is another area where observation may connect to short distance physics and to the geometry of string compactification.  In string models, the potential energy responsible for inflation is closely tied to the dynamics of moduli stabilization.  This sector is generally independent of the sector responsible for Standard Model, though they may couple through supersymmetry breaking.  Just as the Standard Model has many kinds of vacuum energy, so does the moduli sector: branes, fluxes, curvature, curvature, ZPE, instantons, $\alpha'$ effects.  Only in the past few years have these been have these been under enough control to build detailed models.

Slow-roll inflation requires a potential which is a nearly flat function of a scalar field.  One natural mechanism is to have an extra brane-antibrane pair, where the inflaton is the separation.  Models of this type were discussed by Haack~\cite{Haack} and McAllister~\cite{McAllister}.  Another possible potential comes from shape deformations of the compact dimensions (moduli), as discussed by Quevedo~\cite{Quevedo} in large-dimension models.

The phenomenology of inflation centers on two key Cosmic Microwave Background (CMB) parameters, the ratio $r$ of tensor to scalar perturbations and the spectral slope $n_s$.  The default Harrison-Zeldovich model has a scale invariant spectrum, $n_s = 1$, with no tensor modes, $r=0$.  The original slow-roll inflation models made the distinctive predictions of nonzero $r$ and $n_s$ slightly red, though the inflationary model-building industry can now populate much more of the parameter space.

The first detailed string model, KKLMMT, gave $r=0$ and $n_s \sim 0.975$~\cite{KKLMMT}.  The various models discussed at this meeting are shown on Fig.~1.  There is an important distinction between models with $r$ zero or nonzero.  Models with $r=0$ generally involve relatively small changes in the brane positions or moduli, possibly involving only a small region of the compact space, while $r > 0$ requires large changes that generally involve the whole space.
\begin{figure*}[t]
\centering
\includegraphics[width=135mm]{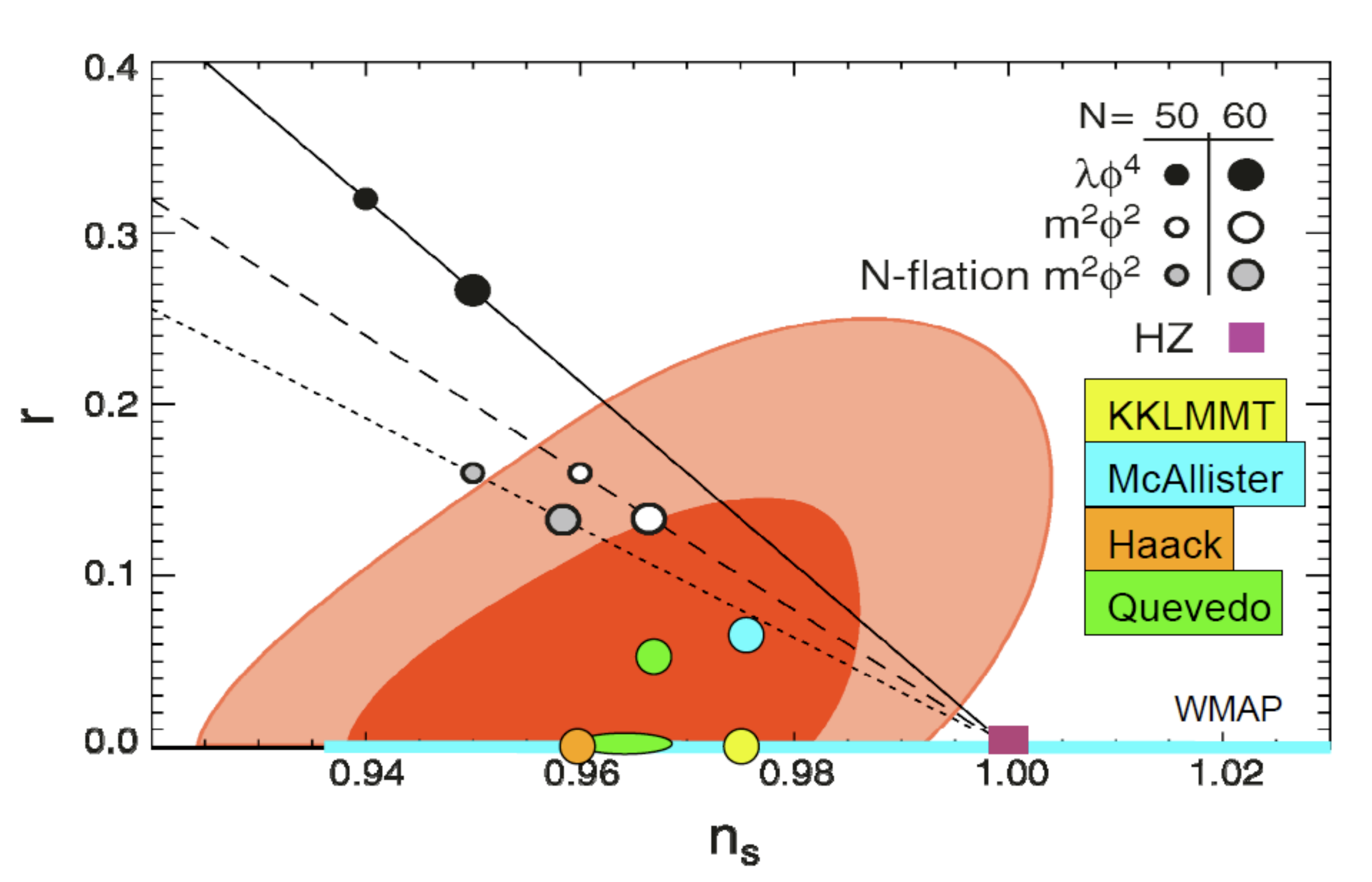}
\caption{Bounds on $r, n_s$ from WMAP5~\cite{WMAP5}.  Standard chaotic inflation models are shown in white and black circles.  Gray and colored circles show a variety of string theory models.} 
\end{figure*}
There will be many new CMB experiments in next five to ten years --- Planck, SPIDER, Clover, QUIET, BICEP, EBEX, PolarBEAR, ... --- and the error bars in both directions will be several times smaller.  While this can't falsify all inflation models, or all string inflation models, it will falsify most of them, and hopefully focus attention on certain kinds of model and compactification.  

Ultimately it would be valuable to have more than two parameters.  Possible additional observables are nongaussianities and cosmic strings.  There have been suggestions that the data give some evidence for these, but at no more than a two-sigma level.

\section{FORMAL THEORY}

Judging from the parallel talks, `formal theory' covers the development of technical tools needed to describe nature, and moreover the application of these tools to a broad range of physics.  The central technical tool, quantum field theory, is around eighty years old, and it is clear that we are still learning new things.  Our new tool, string theory, is not quite half as old.  We have learned some important things about it, notably all the ways in which it is entangled with quantum field theory, and we still have a lot more to learn.

\begin{acknowledgments}
This work was supported in part by NSF grants 
PHY05-51164 and PHY04-56556. 
\end{acknowledgments}


\begin{thebibliography}{9}   

\bibitem{Britto}
R. Britto, these proceedings.

\bibitem{Zanderighi}
G. Zanderighi, these proceedings.

\bibitem{Marquard}
P. Marquard, these proceedings.

\bibitem{Maldacena}
J. Maldacena, these proceedings.

\bibitem{Bern}
Z. Bern, these proceedings.

\bibitem{Rahman}
R. Rahman, these proceedings.

\bibitem{Ward}
B. F. L. Ward, these proceedings.

\bibitem{Rajagopal}
K. Rajagopal, these proceedings.

\bibitem{Myers}
R. Myers, these proceedings.

\bibitem{Starinets}
A. Starinets, these proceedings.

\bibitem{Kachru}
S. Kachru, these proceedings.

\bibitem{Polchinski}
J. Polchinski, ``The cosmological constant and the string landscape,''
in {\it 23rd Solvay Conference In Physics: The Quantum Structure Of Space And Time,}
eds. D. Gross, M. Henneaux, A. Sevrin. Hackensack, World Scientific, 2007. [arXiv:hep-th/0603249].

\bibitem{Straumann}
See N.~Straumann,
  ``The mystery of the cosmic vacuum energy density and the accelerated
  expansion of the universe,''
  Eur.\ J.\ Phys.\  {\bf 20}, 419 (1999)
  [arXiv:astro-ph/9908342].
  
\bibitem{Milton}
K. Milton, these proceedings.

\bibitem{Mannheim}
P. Mannheim, these proceedings.

\bibitem{Nevzorov}
R. Nevzorov, these proceedings.

\bibitem{Gmeiner}
F. Gmeiner, these proceedings.

\bibitem{Ratz}
M. Ratz, these proceedings.

\bibitem{Camara}
P. G. Camara, these proceedings.

\bibitem{Krefl}
D. Krefl, these proceedings.

\bibitem{Weigand}
T. Weigand, these proceedings.

\bibitem{Seiberg}
N. Seiberg, these proceedings.

\bibitem{Haack}
M. Haack, these proceedings.

\bibitem{McAllister}
L. McAllister, these proceedings.

\bibitem{Quevedo}
F. Quevedo, these proceedings.

\bibitem{KKLMMT}
  S.~Kachru, R.~Kallosh, A.~Linde, J.~M.~Maldacena, L.~P.~McAllister and S.~P.~Trivedi,
  ``Towards inflation in string theory,''
  JCAP {\bf 0310}, 013 (2003)
  [arXiv:hep-th/0308055].

\bibitem{WMAP5}
  E.~Komatsu {\it et al.}  [WMAP Collaboration],
  ``Five-Year Wilkinson Microwave Anisotropy Probe (WMAP)
  Observations: Cosmological Interpretation,''
  arXiv:0803.0547 [astro-ph].
\end{thebibliography}
\end{document}